# The Talbot Effect

Masud Mansuripur[†]

The Talbot effect, also referred to as self-imaging or lensless imaging, was originally discovered in the 1830's by Talbot.[1] Over the years, various investigators have found different aspects of this phenomenon, and a theory of the Talbot effect capable of explaining the various observations based on the classical diffraction theory has emerged.[2-4] For a detailed description of the Talbot effect and related phenomena as well as a historical perspective on the subject the reader may consult references 3 and 4 and the references cited therein. Since many of the standard textbooks in optics do not discuss the Talbot effect, I thought it worthwhile to bring the essential features of this phenomenon (and a simple explanation of it) to the attention of the readers of OPN.

**Lensless Imaging of a Periodic Pattern**: The Talbot effect is observed when, under appropriate conditions, a beam of light is reflected from (or transmitted through) a periodic pattern. The pattern may have one-dimensional periodicity (e.g., gratings), or it may exhibit periodicity in two dimensions (e.g., a surface relief structure or a photographic plate imprinted with identical patterns on a regular lattice).

In this section we shall describe the diffraction patterns obtained from a periodic array of cross-shaped apertures. We begin by examining the behavior of an individual aperture under coherent illumination, because the diffraction pattern of a single aperture differs markedly from that of a periodic array of such apertures. Consider the cross-shaped aperture in an opaque screen shown in Fig.1(a). A collimated beam of coherent light (wavelength = $\lambda$) illuminates the screen at normal incidence. The assumed length and height of the aperture are $20\lambda$ each. The logarithmic plots of intensity distribution at distances of $100\lambda$, $200\lambda$, and $600\lambda$ beyond the aperture are computed and shown in Figs.1(b)-(d), respectively.[5] (Note the different scales of these figures.) Beyond $600\lambda$ the intensity distribution will have the far field pattern of Fig.1(d), although its size will scale with distance from the aperture. Under no circumstances do we obtain an intensity pattern that resembles the cross shape of the aperture itself.

Now consider the periodic array of cross-shaped apertures in an opaque screen, shown in Fig.2(a). Each aperture is identical to that in Fig.1(a). The center-to-center spacing between adjacent apertures along both $X$- and $Y$-directions is $p = 60\lambda$. (For simplicity we have assumed the periodic pattern to extend to infinity, although, for practical purposes, a finite number of apertures in a periodic arrangement will suffice.) When the pattern in Fig.2(a) is illuminated with a normally incident, coherent beam of light, the cross shape of the apertures is abundantly reproduced in the intensity patterns obtained at certain distances from the screen. Figures 2(b)-(f) show the computed patterns of intensity distribution at distances $z = 600\lambda$, $1200\lambda$, $1800\lambda$, $2700\lambda$, and $3600\lambda$, respectively.[5] (Note that all pictures in Fig.2 have the same scale.) When the distance from object to image is $z = p^2/\lambda$, as is the case in Fig.2(f), the original pattern of the apertures is reproduced, albeit with a half-period shift in both the $X$- and the $Y$-direction. In Fig.2(d), the distance to the image is $z = p^2/\lambda$, and not only is the original pattern replicated, but also its frequency (along both $X$ and $Y$) has doubled. In Fig.2(c), where the distance to the image is $z = p^2/3\lambda$, the pattern is repeated with three times the original frequency in both $X$ and $Y$ directions. By showing the intensity distribution at other distances from the object, Figs.2(b) and 2(e) emphasize that perfect reproduction of the original pattern does not occur everywhere, but only at certain special planes.

---

[†]Masud Mansuripur is Professor of Optical Sciences at the College of Optical Sciences of the University of Arizona in Tucson. His e-mail address is masud@optics.arizona.edu.

A hint as to why these periodic patterns are repeated at certain intervals may be gleaned from the following argument. A plane wave normally-incident on a periodic structure creates a discrete spectrum of plane waves propagating along the directions

$$\boldsymbol{k} = (k_x, k_y, k_z) = 2\pi\left[m/p, n/p, \sqrt{(1/\lambda)^2 - (m/p)^2 - (n/p)^2}\right]. \quad (1)$$

Now, the $z$-component of this $k$-vector may be approximated as

$$k_z \approx (2\pi/\lambda)[1 - \tfrac{1}{2}(m\lambda/p)^2 - \tfrac{1}{2}(n\lambda/p)^2], \quad (2)$$

provided that $p$ is large enough that, for all $m, n$ values of interest, this Taylor series expansion to first order would suffice. The acquired phase after a propagation distance of $z$ would be

$$k_z z \approx (2\pi z/\lambda) - \pi z(m^2 + n^2)\lambda/p^2. \tag{3}$$

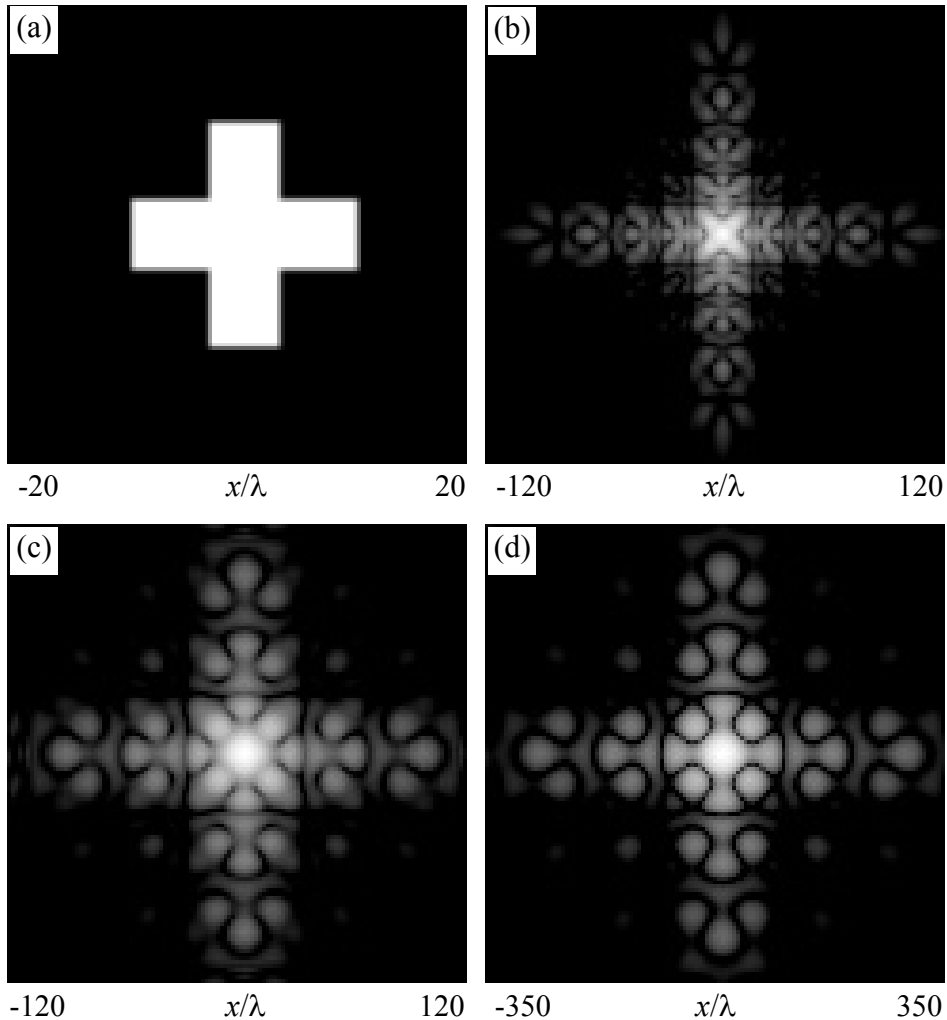

**Figure 1**. (a) A cross-shaped aperture in an opaque screen, illuminated with a normally-incident, monochromatic plane wave of wavelength $\lambda$. The length and the height of the cross are $20\lambda$ each. Also shown are the computed plots of intensity distribution (logarithmic) at various distances $z$ from the aperture: (b) $z = 100\lambda$, (c) $z = 200\lambda$, (d) $z = 600\lambda$. (Note: Scale is different for different pictures.)

Clearly, if $z$ is an even-integer multiple of $p^2/\lambda$, since $m, n$ are also integers, the above phase will differ from the constant $2\pi z/\lambda$ by an inconsequential multiple of $2\pi$. Since all plane-waves emanating from the object will thus arrive at the image plane with the same phase-factor, their superposition will recreate the original pattern.

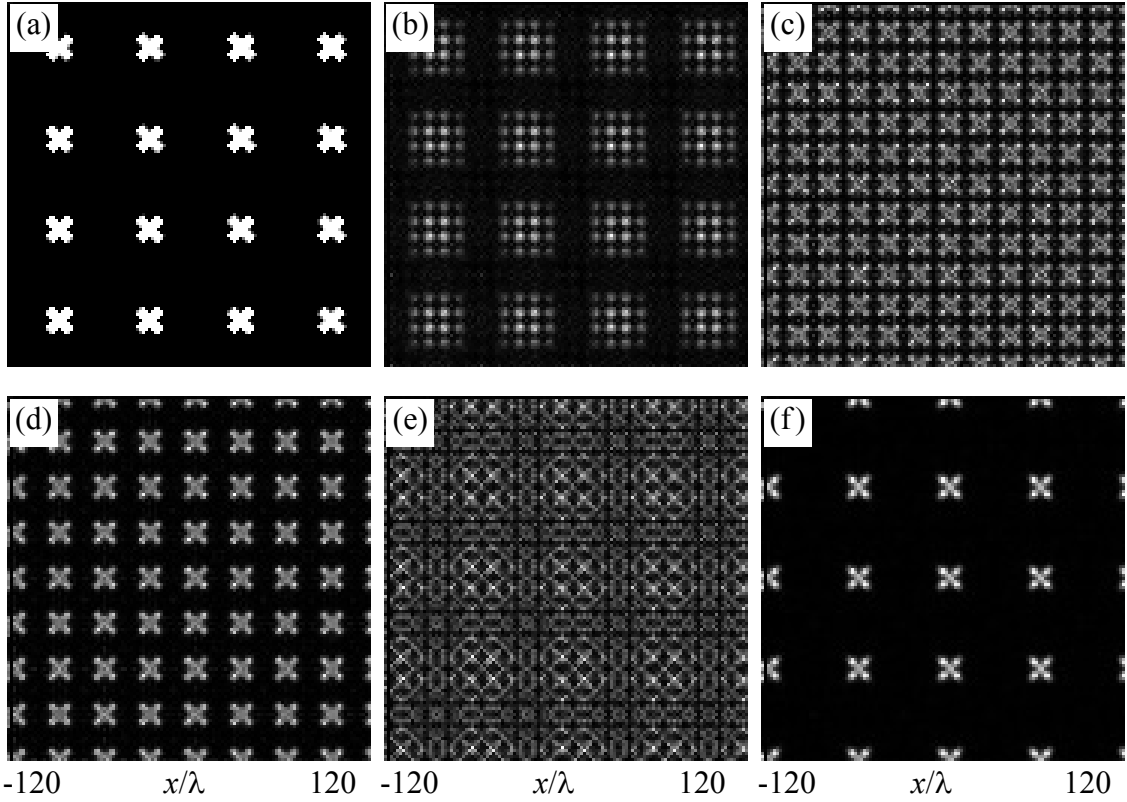

**Figure 2.** (a) Periodic array of cross-shaped apertures in an opaque screen, illuminated with a normally-incident, monochromatic plane-wave of wavelength $\lambda$. The crosses are $20\lambda$-wide on each side. Also shown are the computed plots of intensity distribution at various distances $z$ from the aperture: (b) $z = 600\lambda$, (c) $z = 1200\lambda$, (d) $z = 1800\lambda$, (e) $z = 2700\lambda$, (f) $z = 3600\lambda$. Note that the scale is the same for all the various pictures.

It turns out that $z$ does not need to be an even-integer multiple of $p^2/\lambda$ for self-imaging to occur. At odd-integer multiples of $p^2/\lambda$, for instance, replica of the original pattern will also emerge, but with a half-period shift. Multiple images of the pattern will appear at certain non-integer multiples of $p^2/\lambda$ as well. These aspects of the Talbot effect will be further clarified below, when we present a more rigorous analysis.

Although the mathematical argument that supports the Talbot effect depends on periodicity of the object in the $XY$-plane, certain patterns that are not globally periodic, but appear to be so locally, will also produce self-images. For example, the concentric ring pattern shown in Fig.3, when illuminated by a normally incident, coherent beam, will yield the patterns of Figs.3(b)-(d) at distances $z = 18\lambda$, $27\lambda$, and $36\lambda$, respectively. The period $p$ of the rings is $6\lambda$, and the width of the bright rings is $2\lambda$. Clearly, the self-images break down near the center and near the outer edge, because periodicity is no longer valid in these regions. But a near self-image at $z = p^2/\lambda$ and a frequency-doubled image at $z = p^2/2\lambda$ are clearly observed. Another example is shown in Fig.4, where a spiral pattern with period $p = 9\lambda$ is propagated to distances of $z = 40.5\lambda$, $60.75\lambda$, and $81\lambda$. Again in Figs.4(b) and 4(d) the center and the outer rings are not well reproduced, but nearly everything else is.

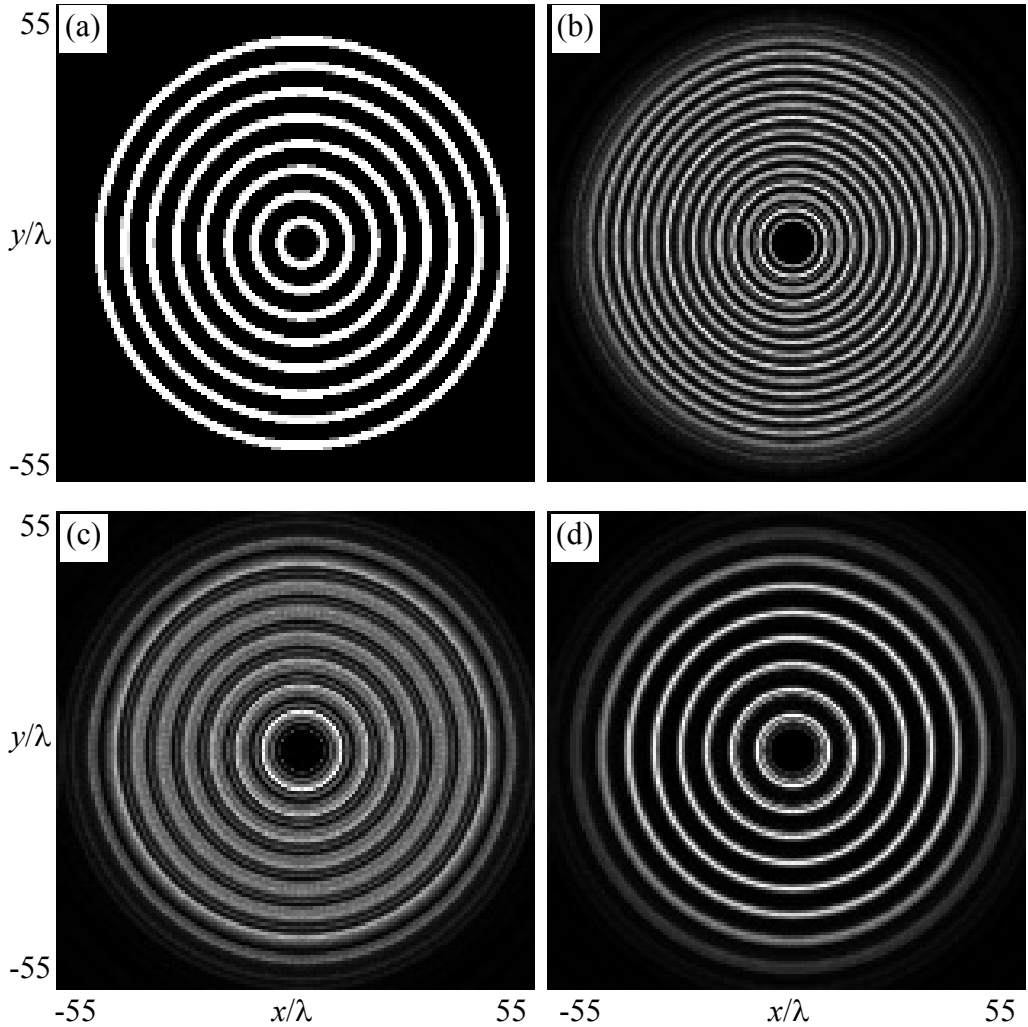

**Figure 3.** (a) A mask consisting of eight concentric rings (width = $2\lambda$, spacing = $6\lambda$) is illuminated with a normally-incident, monochromatic plane-wave (wavelength = $\lambda$). The intensity distributions shown are computed for distances of $18\lambda$ (b), $27\lambda$ (c), and $36\lambda$ (d) from the mask. A bright spike appearing in the central region of each image has been blocked off to improve the image contrast.

The Talbot effect is more general than the above limited exposition may have indicated. The pattern periodicities may be in one- or two-dimensions; the object may modulate both the amplitude and the phase of the light beam; certain applications rely on the use of incoherent light sources; in the case of two-dimensional periodic patterns, the underlying lattice may be square, rectangular, hexagonal, etc.; the incident beam may be a plane wave or a spherical wavefront originating at a point source; applications are not limited to visible light but extend from x-rays to microwaves, to electron and atom optics. To appreciate the variety of arrangements that lead to useful and interesting results the reader is encouraged to consult the published literature.

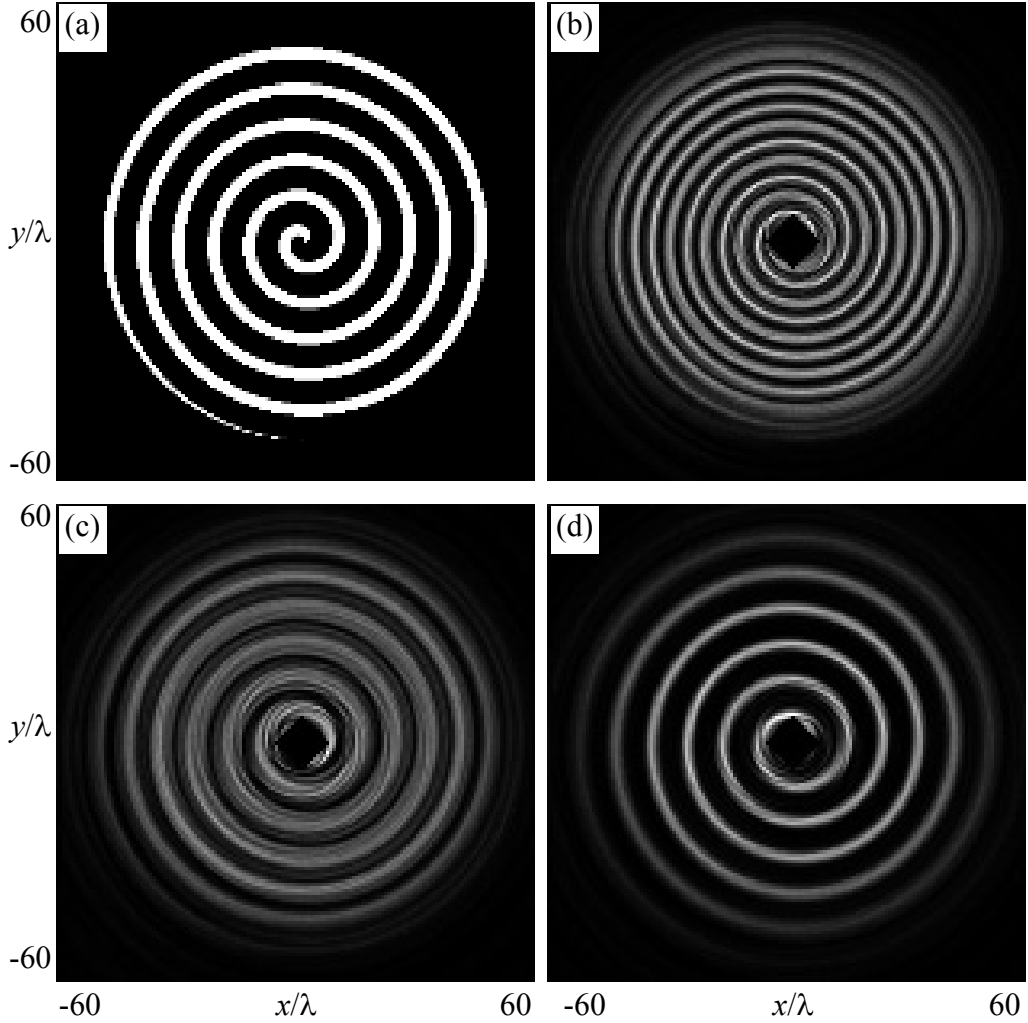

**Figure 4.** (a) A mask consisting of a spiral aperture (width = $3\lambda$, spacing = $9\lambda$) is illuminated with a normally-incident, monochromatic plane wave (wavelength = $\lambda$). The intensity distributions shown are computed for distances of $40.5\lambda$ (b), $60.75\lambda$ (c), and $81\lambda$ (d) from the aperture. A bright spike appearing in the central region of each image has been blocked off in order to improve the image contrast.

**A Simple Analysis**: Consider a point source located at $(x, y, z) = (x_0, y_0, 0)$, radiating a spherical wavefront into the region of space $z > 0$, as shown in Fig.5. In this analysis we assume that all spatial dimensions are normalized by the vacuum wavelength $\lambda$ of the light. Therefore, $\lambda$ will not appear explicitly in any of the following equations. In the $z = z_0$ plane, the complex amplitude distribution may be written

$$A(x, y, z = z_0) = (1/r) \exp(i2\pi r)$$
$$= \exp\left[i2\pi\sqrt{(x - x_0)^2 + (y - y_0)^2 + z_0^2}\right]/\sqrt{(x - x_0)^2 + (y - y_0)^2 + z_0^2}$$
$$\approx (1/z_0) \exp(i2\pi z_0) \times \exp[i\pi(x^2 + y^2)/z_0]$$
$$\times \exp[i\pi(x_0^2 + y_0^2)/z_0] \times \exp[-i2\pi(xx_0 + yy_0)/z_0]. \tag{4}$$

In deriving the above approximate expression we have used, for the exponent, the first term in the Taylor series expansion

$$\sqrt{1+x^2} = 1 + \tfrac{1}{2}x^2 + \cdots. \tag{5}$$

Now, the first two terms on the right-hand side of Eq.(4) are the approximate form of the spherical wavefront emanating from a point source at the origin of the plane $z = 0$. The second term is a constant phase factor which depends on the position $(x_0, y_0)$ of the point source within the $XY$-plane, and the third term is a linear phase factor in $x$ and $y$.

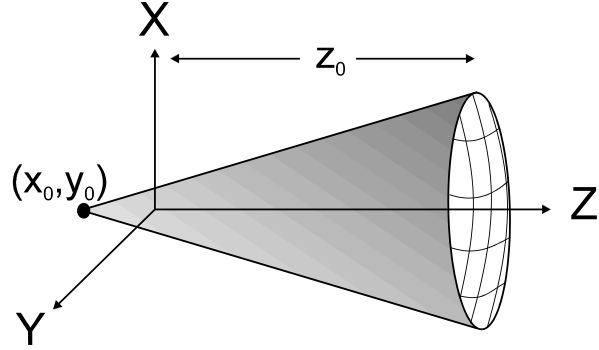

**Figure 5.** A quasi-monochromatic point source located at $(x, y, z) = (x_0, y_0, 0)$ radiates a cone of light into the half-space $z > 0$.

Next, let us assume that a periodic mask is placed in the $XY$-plane at $z = z_0$. Figure 6 shows this situation where the mask has periods $a_x$ and $a_y$ along the $X$ and $Y$ axes. In the general case of phase/amplitude modulation, the complex amplitude transmission function of the mask may be written as follows:

$$t(x, y) = \sum\sum C_{m,n} \exp[i2\pi(mx/a_x + ny/a_y)]. \tag{6}$$

When the incident spherical wavefront is multiplied by $t(x, y)$, each Fourier component of $t(x, y)$ will create a different spherical wavefront which, according to Eq.(4), appears to originate at a different point $(x_0, y_0) = (-mz_0/a_x, -nz_0/a_y)$ in the $XY$-plane. In addition, each such point source appears to have the following phase factor:

$$\exp(i\phi_{mn}) = \exp[-i\pi(x_0^2 + y_0^2)/z_0] = \exp[-i\pi z_0(m^2/a_x^2 + n^2/a_y^2)]. \tag{7}$$

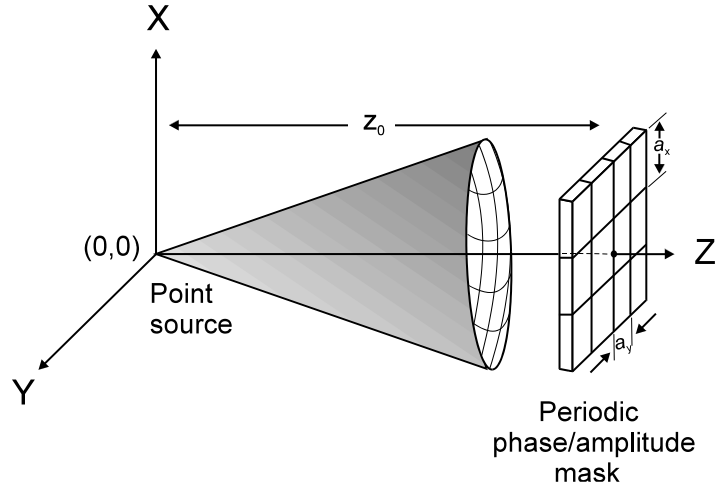

**Figure 6.** A quasi-monochromatic point source located at the origin of the coordinate system illuminates a periodic phase/amplitude mask placed parallel to the $XY$-plane at $z = z_0$. The mask's pattern has period $a_x$ along the $X$-axis and $a_y$ along the $Y$-axis.

The net effect of the grating, therefore, is to replace the single point source with a periodic array of point sources as shown in Fig.7, where the magnitude of each point source is $C_{mn} \exp(i\phi_{mn})$. At the observation plane each point source will give rise to a spherical wavefront which will obey Eq.(4) except for the distance $z_0$ being replaced by $z_0 + z_1$. We thus have

$$A(x,y,z=z_0+z_1) \approx (z_0+z_1)^{-1} \exp[i2\pi(z_0+z_1)] \times \exp[i\pi(x^2+y^2)/(z_0+z_1)]$$
$$\times \sum\sum C_{mn} \exp[-i\pi z_0(m^2/a_x^2 + n^2/a_y^2)] \times \exp[i\pi(m^2/a_x^2 + n^2/a_y^2)z_0^2/(z_0+z_1)]$$
$$\times \exp\{i2\pi[(mz_0/a_x)x + (nz_0/a_y)y]/(z_0+z_1)\}. \tag{8}$$

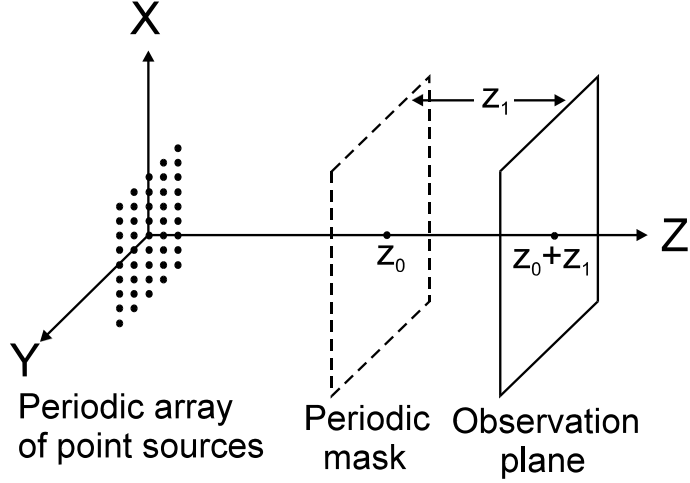

**Figure 7.** Interaction between the periodic mask and the cone of light shown in Fig.6 gives rise to an array of (virtual) point sources, each having a certain phase and amplitude depending on the structure of the mask and its location $z_0$ along the $Z$-axis. To determine the light distribution at the observation plane, one may replace the mask with this "equivalent" array of point sources.

The first term in the above equation is just a spherical wavefront with radius of curvature $z_0 + z_1$. We need not keep track of this term any longer. The last term can be simplified if we define a magnification factor $M = (z_0 + z_1)/z_0$, in which case it is written as

$$\exp\left[i2\pi\left(\frac{mx}{Ma_x} + \frac{ny}{Ma_y}\right)\right]. \tag{9}$$

This is just the plane-wave component $(m,n)$ whose periods $a_x, a_y$ are magnified by a factor $M$. Except for this change of scale, the Fourier coefficients have not changed in going from the plane of the mask $(z = z_0)$ to the observation plane $(z = z_0 + z_1)$. The main factor is the second term in Eq.(8), which can be written as follows:

$$\exp[-i\pi(m^2/a_x^2 + n^2/a_y^2)z_0 z_1/(z_0+z_1)] = \exp[-i\pi(z_1/M)(m^2/a_x^2 + n^2/a_y^2)]. \tag{10}$$

Let us now assume that $a_x^2$ and $a_y^2$ have a "least common multiple" in the following sense:

$$\mu a_x^2 = \nu a_y^2 = a^2, \tag{11}$$

where both $\mu$ and $\nu$ are integers. Then the phase factor in Eq.(10) may be written

$$\exp[-i\pi(z_1/Ma^2)(\mu m^2 + \nu n^2)]. \tag{12}$$

Since $(\mu m^2 + \nu n^2)$ is an integer, if $z_1$ is chosen to be $2\kappa M a^2$ with $\kappa$ being an integer, then the phase factor in Eq.(12) will become unity for all values of *m* and *n*, and can, therefore, be ignored. Under such circumstances Eq.(8) will yield a magnified image of the mask at the observation plane. This is the essence of the Talbot effect.

**Image Multiplicity**: The image multiplicity may be explained for a one-dimensional case, in the special case where the frequency is doubled, but the analysis captures the essence of the phenomenon and can be readily extended to two dimensions and to higher multiplicities. Consider the periodic function $f(x)$ shown in Fig.8(a). Note that the period $a_x$ is much larger than the width of the individual "features" of the function, so that there is plenty of space to insert additional features. Let the Fourier series representation of this function be

$$f(x) = \sum C_m \exp(\mathrm{i}2\pi m x/a_x). \tag{13}$$

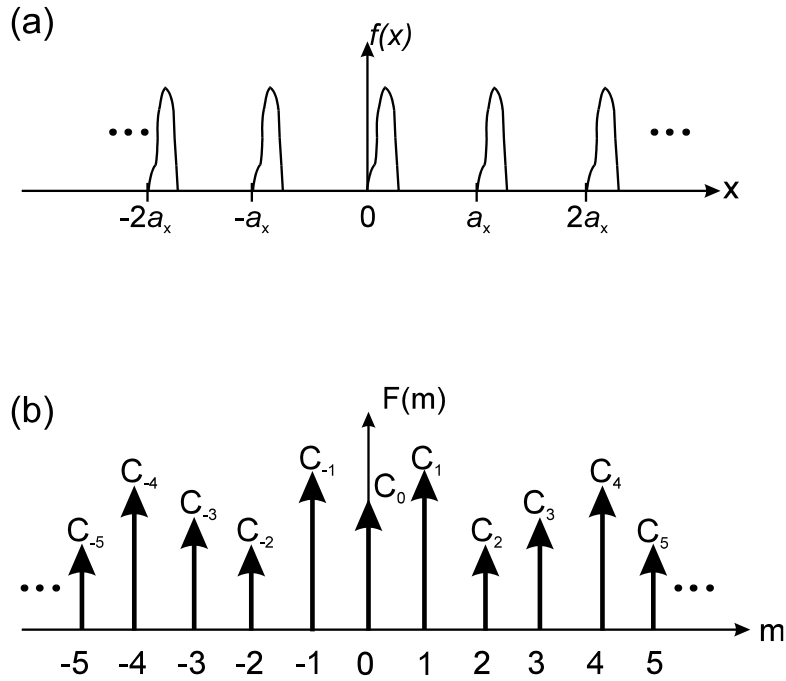

**Figure 8.** (a) Periodic function $f(x)$ in one-dimensional space; individual "features" of the function are much narrower than its period $a_x$. (b) The Fourier transform of $f(x)$ consists of a sequence of $\delta$-functions located at integer multiples of $1/a_x$ in the Fourier domain.

In the Fourier domain, the Fourier transform $F(m)$ of $f(x)$ is a comb function with period $1/a_x$, where the $\delta$-function at position $m$ is multiplied by the corresponding Fourier coefficient $C_m$, as shown in Fig.8(b). Now, let us assume that the odd coefficients of $F(m)$ are multiplied by a complex constant $\beta$. (This would happen in Eq.(12), for instance, if $\mu = 1$, $\nu = 0$, and $z_1 = \tfrac{1}{2}Ma^2$, in which case $\beta = -\mathrm{i}$.) We can then separate the Fourier coefficients of $f(x)$ into even and odd terms, as shown in Fig.9. Both of the resulting comb functions in the Fourier domain have twice the period of the original comb function; therefore, their inverse transforms in the *x*-domain will have twice the frequency. The second comb function in Fig.9 is also shifted by one-half period, which means that its inverse transform must be multiplied by $\exp(\mathrm{i}2\pi x/a_x)$.

The resulting comb functions in the $x$-domain are shown in Fig.10. The net result is that when we add the two comb functions of Fig.10 and convolve the resultant function with the unit-period function $f_0(x)$, we will find the function shown in Fig.11. Because the width of $f_0(x)$ is less than half the period $a_x$, the new features added to the function will not overlap with the old ones, yielding a function with an apparently reduced period. However, the periodicity is only in the amplitude of the function, since the phase of each feature differs from the phase of its adjacent features. In any event, this description explains why the apparent periodicity of the pattern in Fig.2 increased at certain distances.

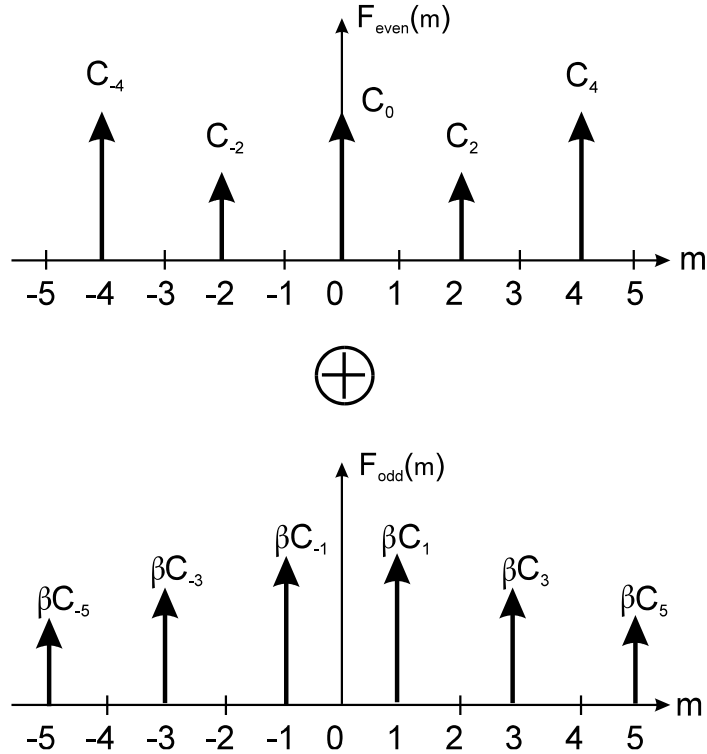

**Figure 9.** In Fig.8(b), when the odd components of the Fourier transformed function $F(m)$ are multiplied by a constant $\beta$, the function may be resolved into two comb functions, $F_{\text{even}}(m)$ and $F_{\text{odd}}(m)$. In these new functions, the spacing between adjacent $\delta$-functions is $2/a_x$ and, in the case of $F_{\text{odd}}(m)$, the function is shifted by half a period.

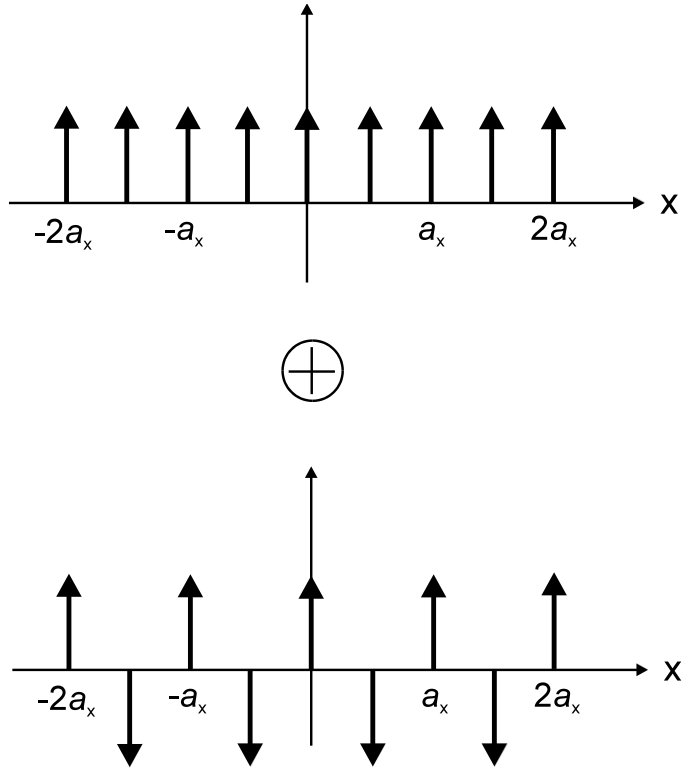

**Figure 10.** The comb function corresponding to $F_{\text{even}}(m)$, when inverse transformed to the $x$-domain, will yield a comb function which has twice the frequency of the original function $f(x)$. Likewise, the inverse transform of the comb function corresponding to $F_{\text{odd}}(m)$ will have a spacing of $a_x/2$ between its adjacent $\delta$-functions, but, because of the half-period shift in the Fourier domain, every other $\delta$-function is flipped.

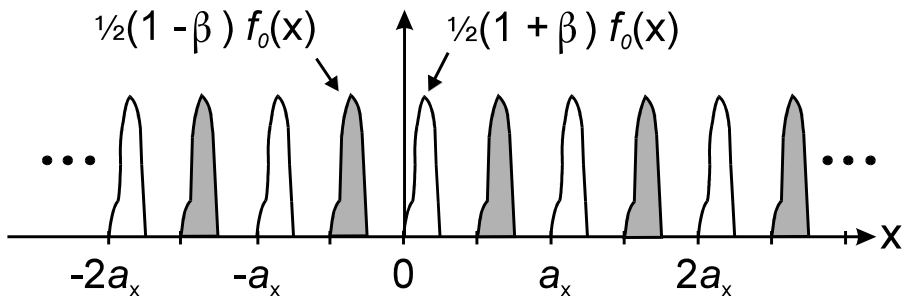

**Figure 11.** When the sum of the two comb functions in Fig.10 is convolved with the individual features $f_0(x)$ of $f(x)$, the resulting function appears to have twice the frequency of the original $f(x)$. Note, however, that the "features" of the new function are alternately multiplied by ½$(1 + \beta)$ and ½$(1 - \beta)$.